\documentstyle[epsfig]{aipproc}
\def\gsim{\;\rlap{\lower 2.5pt
 \hbox{$\sim$}}\raise 1.5pt\hbox{$>$}\;}
\def\lsim{\;\rlap{\lower 2.5pt
   \hbox{$\sim$}}\raise 1.5pt\hbox{$<$}\;}
\newcommand{\beq}{\begin{equation}}
\newcommand{\eeq}{\end{equation}}
\begin{document}
\title{Clusters in the Precision Cosmology Era}

\author{Zolt\'an Haiman$^*$\thanks{Hubble Fellow}, 
Joseph J. Mohr$^{\dagger}$
and Gilbert P. Holder$^{\triangle}$}
\address{$^*$ Princeton University Observatory, Princeton, NJ 08544\\
$^{\dagger}$Departments of Astronomy and Physics, University of 
Illinois, Urbana, IL  61801\\
$^\triangle$ Department of Astronomy and Astrophysics, University of 
Chicago, Chicago, IL 60637}

\maketitle

\begin{abstract}
Over the coming decade, the observational samples available for studies of
cluster abundance evolution will increase from tens to hundreds, or possibly to
thousands, of clusters.  Here we assess the power of future surveys to
determine cosmological parameters.  We quantify the statistical differences
among cosmologies, including the effects of the cosmic equation of state
parameter $w$, in mock cluster catalogs simulating a 12~deg$^{2}$
Sunyaev-Zel'dovich Effect (SZE) survey and a deep 10$^{4}$~deg$^{2}$ X--ray
survey.  The constraints from clusters are complementary to those from studies
of high--redshift Supernovae (SNe), CMB anisotropies, or counts of
high--redshift galaxies.  Our results indicate that a statistical uncertainty
of a few percent on both $\Omega_m$ and $w$ can be reached when cluster surveys
are used in combination with any of these other datasets.
\end{abstract}

\section*{Introduction}

Because of their relative simplicity, galaxy clusters provide a
uniquely useful probe of the fundamental cosmological parameters.  The
formation of the large--scale dark matter potential wells of clusters
is likely independent of complex gas dynamical processes, star
formation, and feedback, and involve only gravitational physics.  The
observed abundance of nearby clusters implies robust constraints on
the amplitude $\sigma_8$ of the power spectrum on cluster scales to an
accuracy of $\sim 25\%$ \cite{white93,viana96}.  In addition, the
redshift--evolution of the observed cluster abundance constrains the
matter density $\Omega_0$ \cite{bahcall98,blanchard98,viana99}.

In order to be useful for these cosmological studies, the masses of
galaxy clusters have to be known.  Existing studies utilized the
presently available tens of clusters with mass estimates
\cite{gioia90,vikhlinin98}, and as a result, were limited in their
scope.  The present samples, however, will likely soon be replaced by
catalogs of thousands of intermediate redshift and hundreds of high
redshift ($z>1$) clusters.  At a minimum, the analysis of the European
Space Agency {\it X--ray Multi--mirror Mission (XMM)} archive for
serendipitously detected clusters will yield hundreds, and perhaps
thousands of new clusters with temperature measurements
\cite{romer00}.  Dedicated X--ray and SZE surveys could likely surpass
the {\it XMM} sample in areal coverage, number of detected clusters or
redshift depth.

The imminent improvement of distant cluster data motivates us to
estimate the cosmological power of future surveys.  In particular, we
study the constraints provided by a 12~deg$^{2}$ SZE survey
\cite{carlstrom99}, or by a deep 10$^{4}$ deg$^{2}$ X--ray survey.
Our primary goals are (1) to quantify the accuracy to which various
cosmological models can be distinguished from a standard $\Lambda$
Cold Dark Matter ($\Lambda$CDM) cosmology; and (2) to contrast
constraints from clusters to those available from CMB anisotropy
measurements, from high--redshift SNe, or from high--redshift galaxy
counts \cite{schmidt98,perlmutter99,newman99}.

The galaxy cluster abundance provides a natural test of models that include a
dark energy component with an equation of state parameter $w\equiv p/\rho \ne
-1$ \cite{turner97,caldwell98,huterer00}.  The value of $w$ directly affects
the linear growth of fluctuations, and the angular diameter distance (and hence
the SZE decrement and the X--ray luminosity) to individual clusters. We
restrict our analysis to a flat universe, and focus on the following four
parameters: the matter density $\Omega_m$; the equation of state parameter $w$
(assumed to be constant); the Hubble constant $h\equiv H_0/100~{\rm
km~s^{-1}~Mpc^{-1}}$' and the amplitude of the power spectrum on $8h^{-1}$Mpc
scales, $\sigma_8$.  A broader range of parameters, including open/closed
universes, and evolving $w(z)$, will be examined in future work
\cite{holder01}. Details of the study described here can be found in
\cite{haiman01}.

\section*{Modeling Details}

\subsection*{General Approach}

To quantify the power of a future cluster survey to distinguish
cosmologies, we utilize the following approach:

\begin{enumerate}

\item We pick a fiducial cosmological ($\Lambda$CDM) model, with
the parameters $(\Omega_\Lambda,\Omega_{\rm
m},h,\sigma_{8},n)=(0.7,0.3,0.65,0.9,1)$, based on present large scale
structure data \cite{bahcall99}. We assume this model describes the
``real'' universe.

\item In the fiducial model, we compute the abundance of clusters
$dN_{\rm fid}/dz$ as a function of redshift in a specific (SZE or
X--ray) survey. This simulates the dataset that will be available in
the future for cosmological tests.

\item We vary parameters of our model, and recompute the cluster
abundance $dN_{\rm test}/dz$ as a function of redshift in this new
``test'' cosmology.  In each test cosmology, we set the value of
$\sigma_8$ by requiring the local cluster abundance at redshift
$z\approx 0$ to match the value in the fiducial cosmology.

\item We compute the likelihood of observing the redshift
distribution $dN_{\rm fid}/dz$ if the true distribution were $dN_{\rm
test}/dz$.  We utilize both the normalization and shape of the
distributions, by combining the standard Poisson and
Kolmogorov--Smirnov tests.

\item We repeat steps 3 and 4 for a wide range of values of $w$, $\Omega_m$, and $h$.

\end{enumerate}

\subsection*{Predicting Cluster Abundance and Evolution}

The fundamental ingredient of this approach is the cluster abundance,
given a cosmology and the parameters of a survey.  In this study, we
utilize the ``universal'' halo mass function found in a series of
recent large--scale cosmological simulations \cite{jenkins00}.
Following these simulations, we assume that the comoving number
density of clusters at redshift $z$ with mass $M$ is given by
\beq
\frac{dn}{dM}(z,M)=
0.315 
\frac{\rho_0}{M}  
\frac{1}{\sigma_M}
\frac{d\sigma_M}{dM}
\exp\left[-\left|0.61-\log(D_z\sigma_M)\right|^{3.8}\right],
\label{eq:dndm}
\eeq
where $\sigma_M$ is the present day r.m.s. density fluctuation on
mass--scale $M$\cite{eisenstein98}, $D_z$ is the linear growth
function, and $\rho_0$ is the present--day mass density. The directly
observable quantity is the number of clusters with mass above $M_{\rm
min}$ at redshift $z\pm dz/2$ in a solid angle $d\Omega$:
\beq
\frac{dN}{dzd\Omega}\left(z\right) =
\left[
\frac{dV}{dzd\Omega}\left(z\right)
\int_{M_{\rm min}(z)}^\infty dM \frac{dn}{dM} 
\right]
\label{eq:dNdzdom}
\eeq 
where $dV/dzd\Omega$ is the cosmological volume element, and $M_{\rm
min}(z)$ is the limiting mass of the survey, as discussed below.
Equations \ref{eq:dndm} and \ref{eq:dNdzdom} reveal that the cluster
abundance depends on cosmology through several quantities: (1) the
growth function $D_z$; (2) the volume element $dV/dzd\Omega$; (3) the
power spectrum $\sigma_M$; (4) the mass density $\rho_0$; and (5) the
survey mass threshold $M_{\rm min}$.  The first four of these
dependencies are well--determined, once the parameters of the
cosmological model and the power spectrum are specified (in
particular, note that the abundance is exponentially sensitive to the
growth function).  The scaling of the limiting mass with cosmology
depends on the specific survey.

\begin{figure}[b!] 
\centerline{\epsfig{file=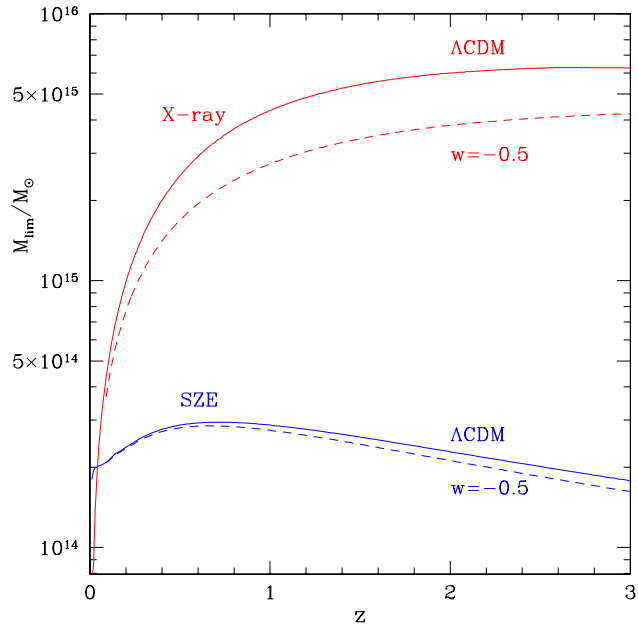,height=3.5in,width=3.5in}}
\vspace{10pt}
\caption{Limiting cluster virial masses for detection in a
10$^4$ deg$^2$ X--ray survey (upper pair of curves) and in a 12
deg$^2$ SZE survey (lower pair of curves).  The solid curves show the
mass limit in our fiducial flat $\Lambda$CDM model, with $w=-1$,
$\Omega_m=0.3$, and $h=0.65$, and the dotted curves show the masses in
the same model except with $w=-0.5$.}
\label{fig:mlim}
\end{figure}

\begin{figure}[b!] 
\centerline{\epsfig{file=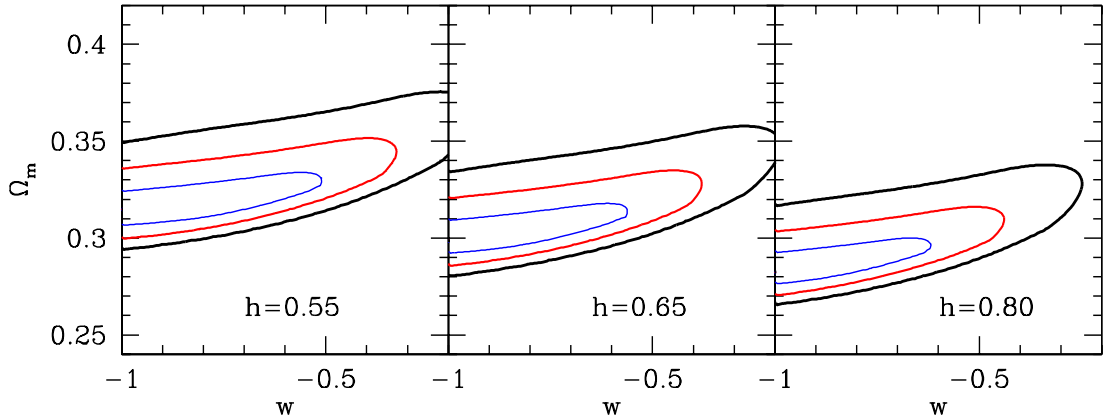,height=2.8in,width=6.3in}}
\vspace{10pt}
\caption{Contours of 1, 2, and 3$\sigma$ likelihood for different
models when they are compared to a fiducial flat $\Lambda$CDM model
with $\Omega_m=0.3$ and $h=0.65$, using the SZE survey.  The three
panels show three different cross--sections of constant total
probability at fixed values of $h$ (0.55,0.65, and 0.80) in the
investigated 3--dimensional $\Omega_m,w,h$ parameter space.}
\label{fig:SZ}
\end{figure}

\begin{figure}[b!] 
\centerline{\epsfig{file=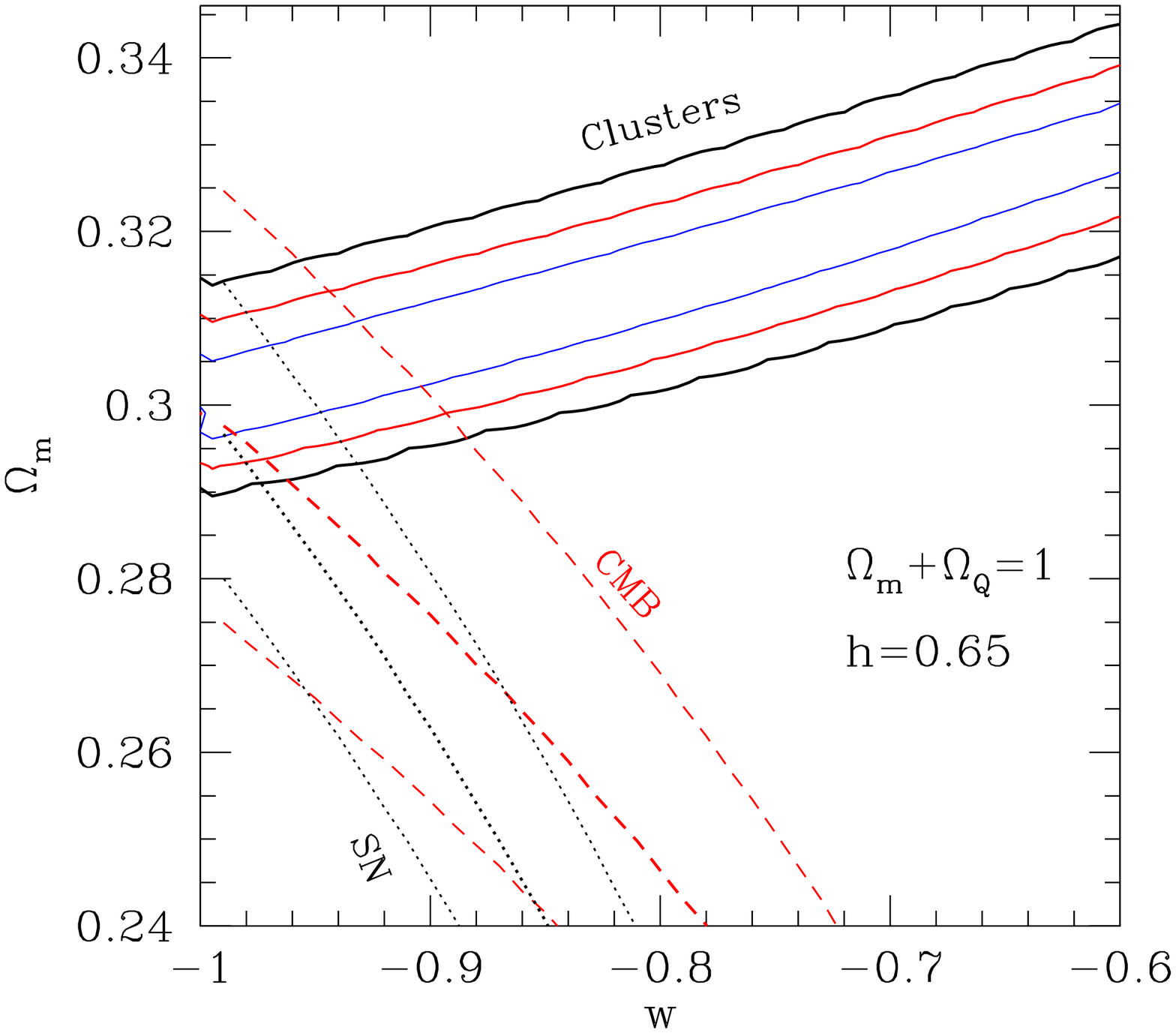,height=3.5in,width=3.5in}}
\vspace{10pt}
\caption{Likelihood contours for a fixed $h=0.65$ in the X--ray survey.  
Also shown are combinations of $w$ and $\Omega_m$ that keep the
spherical harmonic index $\ell$ of the first Doppler peak in the CMB
anisotropy data constant to within $\pm 1\%$ (dashed lines); and
combinations that keep the luminosity distance to redshift $z=1$
constant to the same accuracy.}
\label{fig:XR}
\end{figure}

\subsection*{Cluster Surveys}
\label{subsec:surveys}

In this study, we examine two specific surveys, in which clusters are
detected through either their SZ decrements, or X--ray fluxes. In
practice, the only survey details we utilize in our analyses are the
virial mass of the least massive, detectable cluster (as a function of
redshift and cosmological parameters), and the solid angle of the
survey.  The SZE survey we consider is that proposed by J. Carlstrom
and collaborators \cite{carlstrom99}.  This interferometric survey
will detect clusters more massive than $\sim2\times10^{14}M_\odot$,
nearly independent of their redshift, and will cover an area of
12~deg$^2$ in a year.  The relatively small solid angle of the survey
will allow cluster redshifts to be determined by deep optical and near
infrared followup observations.  The X--ray survey we consider is
similar to a proposed Small Explorer mission, called the Cosmology
Explorer, spear-headed by G. Ricker and D. Lamb.  The survey depth is
$3.6\times10^6$~cm$^2$s at 1.5~keV, and the coverage is 10$^4$~deg$^2$
(approximately half the available unobscured sky).  We focus on
clusters which produce 500 detected source counts in the 0.5:6.0~keV
band, sufficient to reliably estimate the emission weighted mean
temperature.  The X--ray survey could be combined with the Sloan
Digital Sky Survey (SDSS) to obtain redshifts for the clusters.

The most important aspect of both surveys is the limiting halo mass
$M_{\rm min} (z,\Omega_m,w,h)$, and its dependence on redshift and
cosmological parameters.  For an interferometric SZE survey, the
relevant observable is the cluster visibility $V$, which is
proportional to the total SZE flux decrement.  The detection limit as
a function of redshift and cosmology for this survey has been studied
using mock observations of simulated galaxy clusters
\cite{holder00}. In the X--ray survey, $M_{\rm min}$ follows from the
cluster X--ray luminosity -- virial mass relation \cite{bryan98}.
Illustrative examples of the mass limits in both surveys are shown in
Figure~\ref{fig:mlim}, both for $\Lambda$CDM and for a $w=-0.5$
universe.  The SZE mass limit is nearly independent of redshift, and
changes little with cosmology. As a result, the cluster sample can
extend to $z\approx 3$. In comparison, the X--ray mass limit is a
stronger function of $w$, and it rises rapidly with redshift. For the
X--ray survey considered here the number of detected clusters beyond
$z\approx 1$ is negligible.  The total number of clusters in the SZE
survey is $\sim 200$, while in the X--ray survey, it is $\sim 2,000$.

\section*{Results and Conclusions}
\label{sec:results}

Our results are summarized by the likelihood contours in the
$\Omega_m-w$ plane, shown in Figures \ref{fig:SZ} and \ref{fig:XR} for
the SZE and X--ray surveys, respectively.  Figure \ref{fig:SZ} shows
three different cross--sections of constant total probability in the
SZE survey, at fixed values of $h$ (0.55,0.65, and 0.80) in the
investigated 3--dimensional $\Omega_m,w,h$ parameter space.  These
diagrams demonstrate that the constraints on $\Omega_m$ are at the
$\sim 10\%$ level, while $w$ remains essentially unconstrained ($w
\lsim -0.2$).  Nevertheless, the narrowness of the contours in
Figure \ref{fig:SZ} implies that the SZE survey can yield accurate
constraints on $w$ if combined with other data.  We find that the
differences among cosmologies in the SZE survey are driven nearly
entirely by the growth function $D_z$.  This results from the cluster
sample extending to high redshifts ($z>1$), where the growth functions
in different cosmologies diverge rapidly. This makes the SZE sample
especially useful.  For comparison, galaxy counts at $z\sim 1$ probe
mostly the cosmological volume, and constitute and independent test
from clusters \cite{newman99}.  Note that our constraints scale weakly
with $h$: this arises from the weak dependence of the power spectrum
on $h$.

Figure \ref{fig:XR} shows constraints in the X--ray survey for a fixed
$h=0.65$.  The increased number of clusters translates to
significantly narrower contours compared to the SZE survey.  The
orientation of the contours remains similar, but we find that the
cosmological sensitivity arises nearly entirely from the mass limit
$M_{\rm min}$. Indeed, the X--ray flux is more sensitive to cosmology
than the SZE decrement (cf. Fig. \ref{fig:mlim}), and the X--ray
sample extends only out to $z\sim 1$, where the growth functions are
less divergent.  Also shown in Figure \ref{fig:XR} are constraints
expected from CMB anisotropies and from high--$z$ SNe.  The dashed
curves correspond to the CMB constraints (a $\pm 1\%$ determination of
the position of the first Doppler peak); the dotted curves to the
constraints from SNe (a $\pm 1\%$ determination of the luminosity
distance to $z=1$).  As these curves show, the constraints from the
CMB and SNe data are complementary to the direction of the parameter
degeneracy in cluster abundance studies, making the cluster surveys
especially valuable.


Our findings suggest that cluster surveys will lead to tight
constraints on a combination of $\Omega_m$ and $w$, especially
valuable because of their high complementarity to constraints from CMB
anisotropies, magnitudes of high--$z$ SNe, or counts of high--$z$
galaxies.  In combination with either of these data, clusters can
determine both $\Omega_m$ and $w$ to a few percent accuracy.  We have
focused primarily on the statistics of cluster surveys: further work
is needed to clarify the role of systematic uncertainties, arising
from the cosmology--scaling of the mass limits and the the cluster
mass function, as well as our neglect of issues such as galaxy
formation in the lowest mass clusters.

\vspace{\baselineskip}
We thank J. Carlstrom and the COSMEX team for providing access to
proposed instrument characteristics.  ZH acknowledges support from
a Hubble Fellowship.

\end{document}